\documentclass[twocolumn]{aastex62}

\usepackage{amsmath}
\usepackage{CJKutf8}
\received{\today}
\submitjournal{\apjs}

\begin{document}
\begin{CJK}{UTF8}{gbsn}
\title{The measurement of masses of OB-type stars from LAMOST DR5 }

\author{Zhenyan Huo}
\affiliation{Department of Physics, Hebei Normal University, Shijiazhuang 050024, People's Republic of China}

\author[0000-0002-0786-7307]{Zhicun Liu}
\altaffiliation{Hebei Normal University postdoctor}
\affiliation{Department of Physics, Hebei Normal University, Shijiazhuang 050024, People's Republic of China}
\affil{Co-first author}

\author[0000-0003-1359-9908]{Wenyuan Cui}
\affiliation{Department of Physics, Hebei Normal University, Shijiazhuang 050024, People's Republic of China}

\author[0000-0002-1802-6917]{Chao Liu}
\affiliation{Key Laboratory of Space Astronomy and Technology, National Astronomical Observatories, Chinese Academy of Sciences, Beijing 100101, People's Republic of
China}
\affiliation{Institute for Frontiers in Astronomy and Astrophysics, Beijing Normal University, Beijing 102206, Peopleʼs Republic of China}

\author[0000-0002-4828-0326]{Jiaming Liu}
\affiliation{Department of Physics, Hebei Normal University, Shijiazhuang 050024, People's Republic of China}

\author{Mingxu Sun}
\affiliation{Department of Physics, Hebei Normal University, Shijiazhuang 050024, People's Republic of China}

\author{Shuai Feng}
\affiliation{Department of Physics, Hebei Normal University, Shijiazhuang 050024, People's Republic of China}

\author{Linlin Li}
\affiliation{Department of Physics, Hebei Normal University, Shijiazhuang 050024, People's Republic of China}

\correspondingauthor{Wenyuan Cui and Chao Liu}
\email{wenyuancui@126.com;liuchao@nao.cas.cn}
 
\begin{abstract}

The measurements of masses and luminosities of massive stars play an important role in understanding the formation and evolution of their host galaxies. In this work, we present the measurement of masses and luminosities of 2,946 OB-type stars, including 78 O-type stars and 2,868 B-type stars, based on their stellar parameters (effective temperature, surface gravity, and metallicity) and PARSEC isochrones model. Our results show that the median mass and luminosity of the 2,946 OB-type stars are 5.4\,M$_{\sun}$ and \,log\,(L/$\rm {L_\sun}$)=3.2 with the median relative error of 21.4{\%} and 71.1{\%}, respectively. A good agreement between our results estimated by using our method and those derived by using the orbital motions of binary stars from the literature is found for some B-type stars. In addition, we also fit the mass-luminosity relation of B-type stars by using our derived mass and the luminosity from $Gaia$ DR3. 

\end{abstract}

\section{Introduction} \label{Intro}

OB-type stars, which have spectral types of O and B, are young, bright massive stars and have an crucial effect on the star formation and evolution of their host galaxies via their stellar winds and intense radiation \citep{2009ssc..book.....G,2011A&A...530A.108E,2012A&A...539A.143N}. Although the survey of massive stars in the solar neighborhood is less than small-mass stars due to the limit of observation and star formation history, the research of their stellar parameters and chemical composition can help us better understand the formation and evolution of the Milky Way \citep{1992A&AS...94..569L,2006A&A...446..279C,2019A&A...625A.120B}. 

Estimating stellar masses is a fundamental process that improves the understanding of stellar evolutionary processes \citep{Chevalier2023}. The study of stellar masses and ages has an significent impact on many astrophysical aspects, such as the initial mass function (IMF) of a galaxy, the mass-luminosity relation (MLR), and the stellar evolution path \citep{2003PASP..115..763C,2020ARA&A..58..577S,2023Natur.613..460L}. The IMF, a fundamental parameter of stellar population, describes the proportion of stars of different masses at the time of star formation. \citep{Lamb2012}. At the beginning of the 20th century, the mass-luminosity relation, one the most famous empirical ``laws" is a vital way to obtain stellar masses \citep{Henry2004, Henry1993, Xia2010}. The stellar initial masses determine the evolution path of their host galaxies \citep{2009ssc..book.....G}.

\begin{figure*}[ht]
    \centering
    \includegraphics[width=18cm,height=5cm]{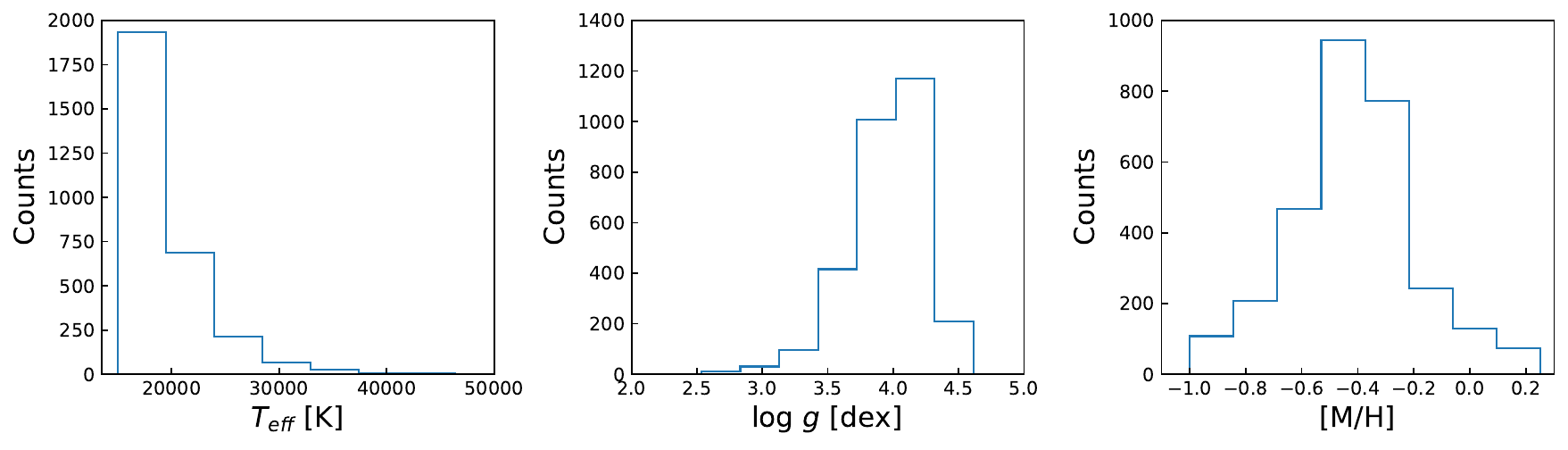}
    \caption{The histograms show the distribution of 2,946 OB-type stars in effective temperature (left panel), surface gravity (middle panel), and metallicity (right panel).}\label{fig:1}
\end{figure*}

The stellar masses are usually estimated by interpolating their stellar parameters (effective temperature ($T_{\rm eff}$), surface gravity (log\,$g$) and metallicity ([M/H])) in the Hertzsprun-Russell diagram (HRD) or the theoretical isochrones. \citet{Hohle2010} derived masses and ages of 2,398 massive stars, which including O-type stars, early B-type stars, and red supergiants, by using the theoretical isochrones. The masses of 1,000 FGK-type single stars are obtained by \citet{Pinheiro2014}, based on the stellar evolution models, MLR, and surface gravity spectroscopic observations, respectively, and their results show good consistency. \citet{Zhang2019} also obtained the masses, radii, and ages of 1,320 binaries by comparing their stellar atmospheric parameters and the stellar evolution model. In addition, \citet{Takeda2007} and \citet{Pont2004} argued that interpolation between isochrones could not explain either the nonlinear mapping of time on the HRD or the nonuniform distribution of stellar masses observed in galaxies, because interpolation is a method of extrapolation, which may generate absurd results.

However, it is difficult to estimate the stellar ages directly from observations or the fundamental laws of physics. The stellar ages are usually inferred indirectly from photometry and spectral observation combined with stellar evolution models \citep{Soderblom2010}. Astroseismology can provide stellar masses and ages with uncertainties in the range of 10-20{\%} \citep{Gai2011, Chaplin2014}. However, the method is only applicable to a limited number of stars with sufficiently and accurately high-resolution photometric parameters. Therefore, one of the most useful methods for deriving the ages of large stellar samples is to match the stellar evolution models under given accurate atmospheric parameters. The study of stellar masses, luminosities, and radii is very important for us to understand the formation and evolution of the Milky Way.

In this work, we aim to determine the masses of OB-type stars from LAMOST DR5. The data of sample is described in Section \ref{Data}; Section \ref{Methods} introduces the method for estimating the stellar masses and luminosities; We present the results and discussion for this work in Section \ref{Results}. Finally, we give a conclusion in Section \ref{Conclusions}.

\section{DATA} \label{Data}

\subsection{The LAMOST data}

The Large Sky Area Multi-Object Fiber Spectroscopic Telescope (LAMOST), which is also called the Guo Shou Jing telescope, innovatively presents the world's unique and largest aperture large field telescope. The vast majority of LAMOST spectra come from objects in the Milky Way. Since 2011, the LAMOST survey has released more than 20 million astronomical spectra covering the wavelength range of 369-910\, nm with a resolution of 1800, building the world's largest database of the astronomical spectra \citep{Cui2012, Zhao2012, Deng2012}. 

\subsection{Sample Selection}

We use the OB catalog of \citet{Liu2019}, which contains 16,032 OB-type stars selected from LAMOST DR5 based on the stellar spectral line indices, as our initial sample data. \citet{Guo2021} used the data-driven technique Stellar LAbel Machine (SLAM) to derive their stellar atmospheric parameters ($T_{\rm eff}$, log\,$g$, [M/H] and $v$\,sin\,$i$), based on the non-LTE TLUSTY synthetic spectra.

Considering the results of extrapolation and the uncertainties of stellar parameters from \citet{Guo2021}, the effect from the spectral signal-to-noise of g-band (S/N)$_g$, and the limitation of the TLUSTY model grids, we used the following criteria to select sample further:

\begin{equation}\label{equation1}
\left\{
    \begin{aligned}
       \rm Remove\,Oe/Be\,stars\\
        \rm (S/N)_g \geq 40 \\
         T_{\rm eff} \geq 15000\,K \\ 
        \rm 1.75\,dex \leq log\,g \leq 4.75\,dex \\
        \rm -1.0 \leq [M/H] \leq 0.3\\ 
    \end{aligned}
\right.    
\end{equation}

Applying the above criteria (\ref{equation1}), we obtain 2,946 normal OB-type stars including 78 O-type stars and 2,868 B-type stars \footnote{We define the massive OB-type stars as the ``normal" OB-type stars, and our sample does not contain 57 OB-type stars, which are not within PARSEC model, and special OB-type stars, such as: sdOB, Oe/Be, WD.}. Figure \ref{fig:1} shows the distribution of the stellar parameters of 2,946 OB-type stars, we can see that most of the stars have surface gravity values (log\,$g$ =4.0\,dex) of main-sequence stars and lower metallicity than the Sun. In other words, the majority of our selected sample are the main-sequence OB-type stars.

\section{Method}\label{Methods}

\subsection{Evolutionary models: PARSEC}

PAdova and TRieste Stellar Evolution Code (PARSEC), which is widely used by the astronomical community \citep{Bressan1993, Girardi2012}, is the first applied by \citet{Bressan2012}, and \citet{Chen2014, Chen2015, Tang2014} has extended this model. The PARSEC model covers the metallicity of -2.2 $\leq$ [M/H] $\leq$ +0.5, the mass range 0.1\,M$_{\sun}$ $\leq$ M $\leq$ 150\,M$_{\sun}$ and a broad age distribution from pre-main sequence to 30\,Gyr very low-mass stars. The feature of the PARSEC model can make us determine the masses and luminosities of the OB-type stars.

\subsection{Determination of masses of OB-type stars}

We use the stellar parameters ($T_{\rm eff}$, log\,$g$ and [M/H]) of 2,946 OB-type stars, which is from \citet{Guo2021}, to interpolate the PARSEC model, and determine their masses. Considering the stellar parameters of OB-type stars derived using  their LAMOST low-resolution spectra (R\,$\sim$1800), we use the uncertainties of atmosphere parameters ($\Delta$$T_{\rm eff}$=1600\,K, $\Delta$log\,$g$=0.25\,dex) and metallicity of \citet{Guo2021}. The uncertainties of stellar masses are also obtained by applying Monte Carlo sampling by assuming a Gaussian distribution with the uncertainties of stellar parameters ($T_{\rm eff}$, log\,$g$ and [M/H]). The final results are listed in Table \ref{chartable}. 
\citet{Rolleston2000} used about 80 B-type main-sequence stars in Galactic open clusters$/$associations to obtain the metallicity gradient of -0.07 dex/kpc. Ideally, the metallicity gradient distribution is -0.3$<$[M/H]$<$+0.2 for our OB-type stars with galactocentric distances of approx 5.5-12.2\,kpc.  However, most of our OB-type stars are at located in galactocentric distances of 9-12\,kpc. Moreover, there may be a systematic difference between the metallicity of \citet{Rolleston2000} and the statistical metallicity of \citet{Guo2021}. Taking into account the above conditions, we also obtain the masses of OB-type stars at solar metallicity and expressed as $M_{2}$ in Table \ref{chartable}.

To verify the effect of the degeneracy of stellar parameters on stellar masses, we apply Monte Carlo simulation (MCMC, \cite{Foreman-Mackey2013}) to examine the potential distribution of stellar masses. Figure \ref{fig:10} in the section appendix shows the distribution of mass of a B-type star by assuming a Gaussian distribution of its stellar parameters ($T_{\rm eff}$, log\,$g$ and [M/H]), and we can see that it can give a relatively reliable mass by using MCMC due to the relatively concentrated distribution of stellar mass. 

\begin{figure}
    \centering
    \includegraphics[width=8cm,height=7cm]{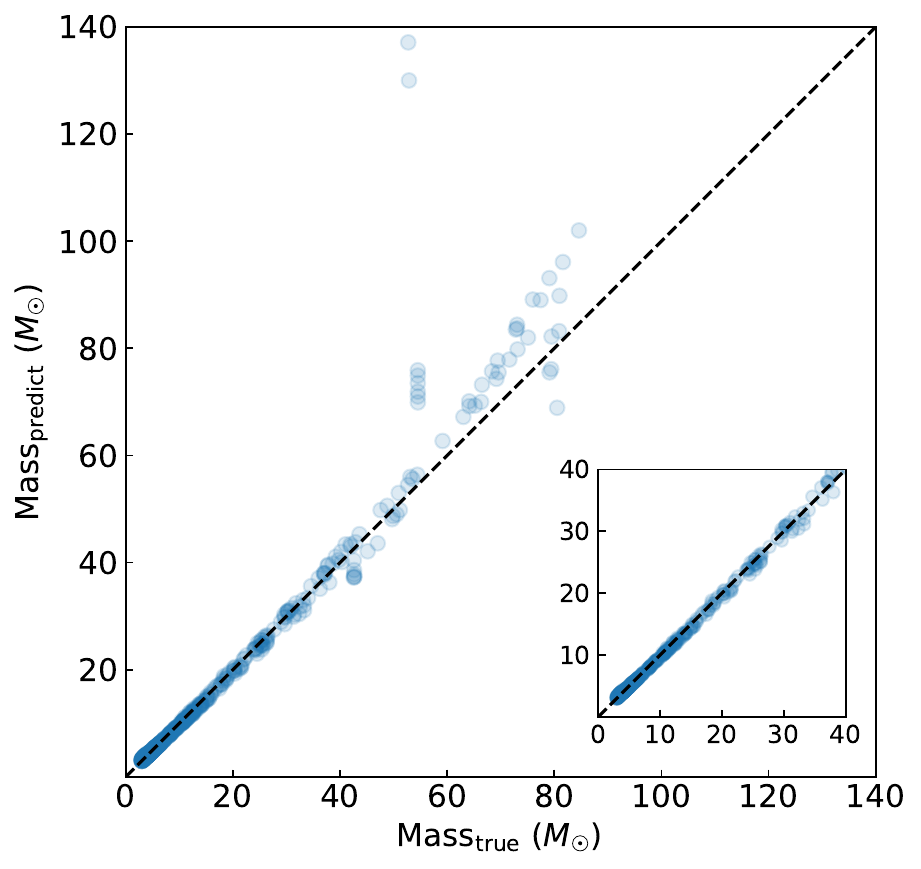}
    \caption{Comparison of the masses from the grid point of PARESEC (Mass$_{\rm True}$) and the masses estimated using our method (Mass$_{\rm predict}$), and the black dashed line represents the one-to-one relation. The subgraph shows the comparison of mass less than 40\,M$_{\sun}$.}
    \label{fig:2}
\end{figure}

\subsection{Test the reliability of the method}\label{subsec:figures}

To inspect the reliability of the method obtaining stellar masses, we randomly select 600 grid points with different $T_{\rm eff}$, log\,$g$, [M/H], masses from the evolutionary model. The masses of 600 grid points are estimated by using the above method again, based on the typical uncertainties of their stellar parameters (1600\,K for $T_{\rm eff}$, 0.25\,dex for log\,$g$ and 0.1\,dex for [M/H]).

As shown in Figure \ref{fig:2}, for the OB-type stars with masses less than 40\,M$_{\sun}$, there is a good consistency between the masses calculated by the theoretical model and the masses obtained by using our method. For some OB-type supergiants with masses larger than 40\,M$_{\sun}$, our results give larger masses than those from the theoretical model because they are located at the edge of the PARSEC grid. Considering our sample only contains 6 OB-type stars with masses larger than 40\,M$_{\sun}$, our method does not have an effect on our current results, and that also indicates that our method is self-consistent.

\section{Results and discussion}\label{Results}

\begin{figure*}[htp]
    \centering
    \includegraphics[width=12cm,height=10cm]{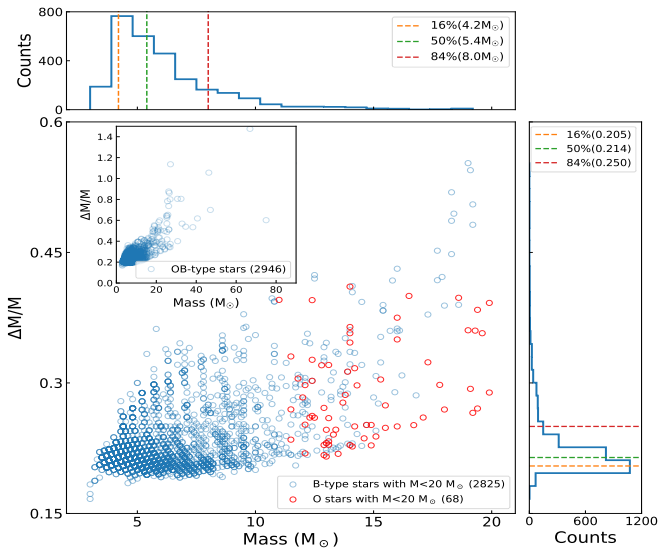}
    \caption{Distribution of 2,893 OB-type stars with masses less than 20\,M$_{\sun}$ in mass (M) vs mass relative errors ($\Delta$\,M/M) diagram. The open blue circle corresponds to the B-type stars and the open red circle to the O-type stars. The subgraph shows the distribution of all OB-type stars in this work. In the top and right histograms, the orange, green, and red dashed lines represent 16{\%}, 50{\%}, and 84{\%} of mass and the relative error of mass, respectively.}
    \label{fig:3}
\end{figure*}

\begin{figure}[htp]
    \centering
    \includegraphics[width=9cm,height=7cm]{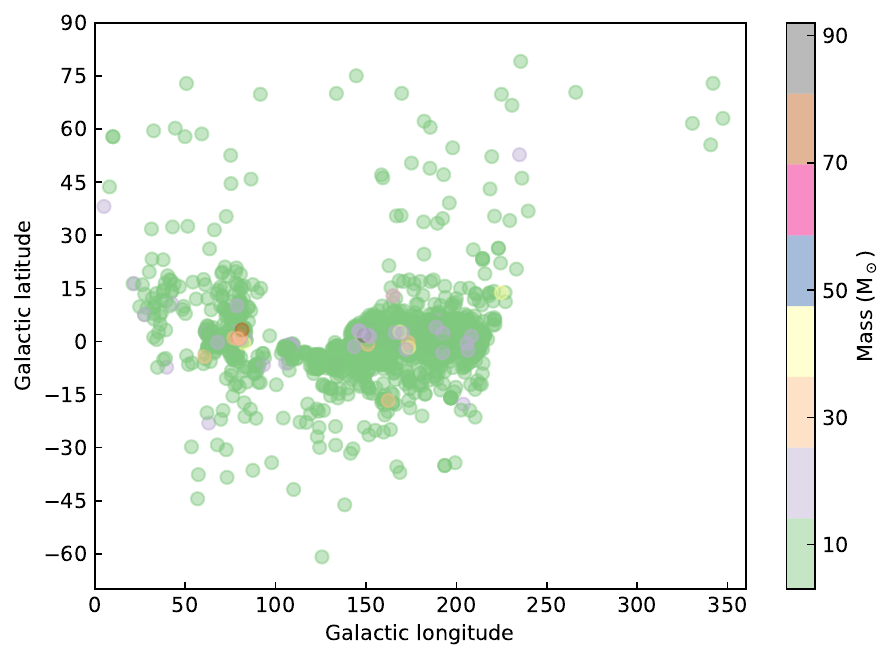}
    \caption{The distribution of OB-type stars in Galactic longitude versus Galactic latitude, and the color code represents the masses of OB-type stars.}
    \label{fig:4}
\end{figure}

\begin{figure}[htp]
    \centering    \includegraphics[width=7.5cm,height=7cm]{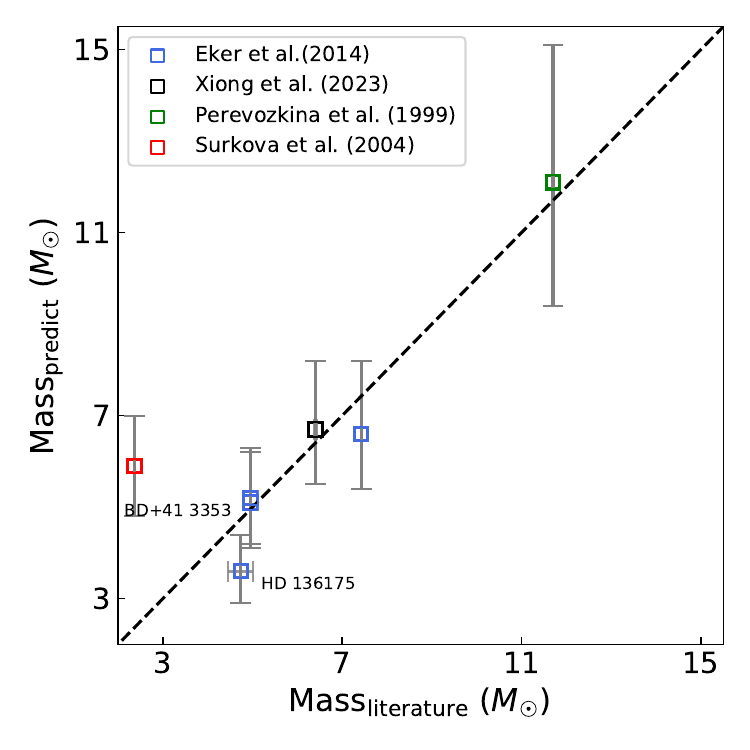}
    \caption{Comparison of the mass estimated in this work and those from the literature, and the black dashed line represents the one-to-one relation.}
    \label{fig:5}
\end{figure}

\subsection{The distribution of Masses}

\begin{figure*}[htp]
    \centering
    \includegraphics[width=16cm,height=7cm]{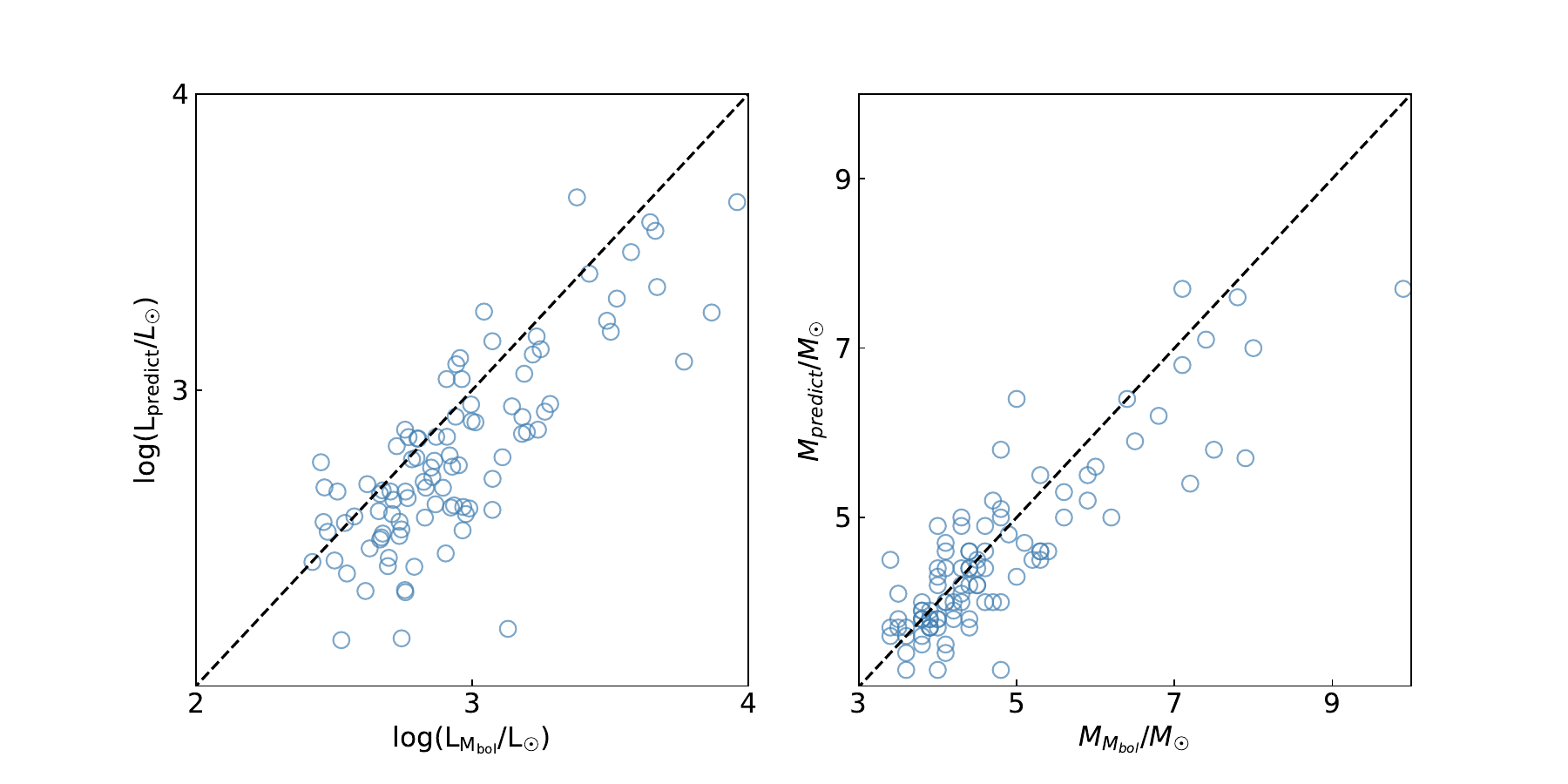}
    \caption{Comparison of luminosities (left panel) and masses (right panel) of 100 OB-type stars derived using Gaia DR3 data to the predicted values obtained by using their stellar parameters and PARSEC model. log(L$_{\rm predicted}$$/$L$_\sun$)) and M$_{\rm predicted}$$/$M$_\sun$ represent the luminosities and masses of OB-type stars obtained using the stellar parameters and PARSEC model, and log(L$_{\rm Mbol}$$/$L$_\sun$)) and M$_{\rm Mbol}$$/$M$_\sun$ represent the luminosities and masses of OB-type stars obtained using the $Gaia$ DR3 data (L$_{\rm Mbol}$\,=\,4$\pi$\, R$^2$\,$\sigma$T$^4$ and $g$\,=\,GM$_{\rm Mbol}$/R$^2$). The dashed lines represent the one-to-one correlation.}
    \label{fig:6}
\end{figure*}

\begin{figure*}[htp]
    \centering
    \includegraphics[width=12cm,height=10cm]{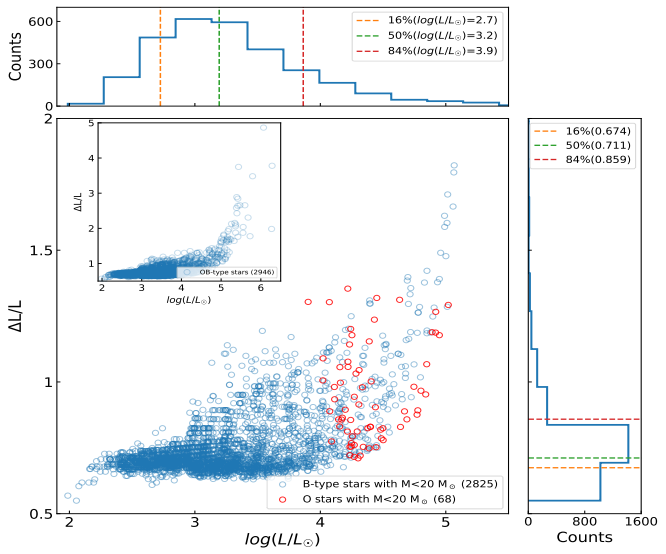}
    \caption{Similar to Figure \ref{fig:3} for the estimated luminosities and relative errors of OB-type stars in this work.}
    \label{fig:7}
\end{figure*}

Figure \ref{fig:3} shows the distribution of 2,893 OB-type stars with masses \footnote{All the masses involved inside our hereafter discussion are M.} less than 20\,M$_{\sun}$, including 2,825 B-type stars and 68 O-type stars, in mass versus relative error of mass ($\Delta$\,M/M) panel, and O-type stars have larger masses than B-type stars. From the top histogram, it is seen that the masses of 84\,$\%$ OB-type stars are less than 8.0\,M$_{\sun}$, the mean value of the masses is about 5.4\,M$_{\sun}$, the masses of 16\,$\%$ OB-type stars are less than 4.2\,M$_{\sun}$, and the stellar masses range are mainly concentrated in 4-10\,M$_{\sun}$ with the peak distribution in 5-6\,M$_{\sun}$. This is due to our sample selection effect that the stars are mainly distributed in the range of $T_{\rm eff}$\,$<$\,20\,kK and 3.5\,$<$\,log\,$g$\,$<$\,4.5\,dex (see Figure \ref{fig:1}). From the right histogram, it can see that $\Delta$\,M/M of 84\,$\%$ OB-type stars is also less than 25\,$\%$. Moreover, some OB-type stars with stellar winds, such as B-type supergiants and O-type stars, are not suitable to use the plane-parallel Tlusty model to derive their stellar parameters, which require spherical model atmospheres (e.g., FASTWIND, CMFGEN, PoWR) \citep{Hawcroft2021, 2006A&A...446..279C}. Thus, the relative errors in the parameters of such kind of stars are large and less accurate, and we also marked these stars in Table \ref{chartable}. The subgraph of Figure \ref{fig:3}, which contains all 2,946 OB-type stars, shows that the stars with masses larger than 20\,M$_{\sun}$ are relatively few. This is also consistent with the distribution of stellar effective temperature. 

The majority of massive stars are members of binary systems so the majority of LAMOST targets will probably be binaries, which maybe impact on $T_{\rm eff}$, log\,$g$\, and luminosity inferred for each star \citep{Sana2012,Moe2017}. However, the results of \citet{Guo2021} indicate that the predicted atmospheric parameters of early-type stars are in good agreement with the results obtained by using their high-resolution spectra within the errors, which imply that the binary have no obvious effect on the stellar parameters of OB-type stars of \citet{Guo2021}.

Figure \ref{fig:4} shows the spatial distribution of 2,946 OB-type stars. The color code represents the masses of the stars. The OB-type stars with larger masses are mainly located in the low Galactic latitude region, which is consistent with the distribution of the Galactic neutral atomic hydrogen (\ion{H}{1}) column density \citep{2016A&A...594A.116H}.

\subsection{Comparison with other works: Mass}

In general, the solution of the orbital motion of binary stars is a reliable method for determining stellar masses \citep{Torres2010}. To validate the reliability of mass, we cross-correlated the binary mass sample of \citet{Xiong2022}, the dynamical mass sample of \citet{Eker2014}, \citet{Perevozkina1996} and \citet{Surkova2004}, respectively. 
Figure \ref{fig:5} shows the comparison of our results and their results for 7 common stars. We can see that most of the stars give a good agreement, except two stars HD\,136175 and BD+41 3353, and we find that the different masses of these two stars are due to the difference between our stellar parameters and those of the literature (atmospheric parameters from photometric data
). This also supports the reliability of our estimated masses for the OB-type stars. 

In addition, we also note that \citet{Hohle2010} also provides a mass sample of OB-type stars by matching the stellar theoretical models, which is similar to the method adopted in this work. By cross-matching the coordinates, out of these 2946 objects, we have 9 objects in common with \citet{Hohle2010}. The isochronous models and stellar atmosphere parameters utilized in \citet{Hohle2010} differ from those used in this work, which will result in a significant discrepancy in the masses for the same OB-type star. We do not use their results as the test sample as a result.

To verify the reliability of masses of OB-type stars obtained by using their stellar parameters, we cross-matching our data with the $Gaia$ DR3 data and randomly select 100 OB-type stars with parallax$\_$error$/$parallax$\leq$\,20\,$\%$, RUWE$<$\,1.4, and \\distances$<$\,4\,kpc. We use the $Gaia$ DR3 luminosities of 100 OB-type stars to calculate their masses (L\,=\,4$\pi$\, R$^2$\,$\sigma$T$^4$ and $g$\,=\,GM/R$^2$). The left panel of Figure \ref{fig:6} shows a comparison between the luminosities predicted by using the stellar parameters for 100 OB-type stars and the luminosities of $Gaia$ DR3, and it is seen that the luminosities of $Gaia$ DR3 are generally larger than our predicted luminosities, which suggest that the stellar luminosities of $Gaia$ DR3 may be affected by binary systems. The right panel of Figure \ref{fig:6} shows a comparison between the masses obtained by using the PARSEC model for 100 OB-type stars and the masses obtained using their $Gaia$ DR3 data. It can be seen that when the stellar mass is larger than 5\,M$_{\sun}$, the mass obtained from the stellar luminosity is biased towards the mass obtained from the stellar atmospheric parameters, which is because more massive stars have a greater binary ratio \citep{Sana2012}. Therefore, our method can obtain the masses of more OB-type stars with less influence from binary stars.

\begin{figure*}
    \centering
    \includegraphics[width=12cm,height=10cm]{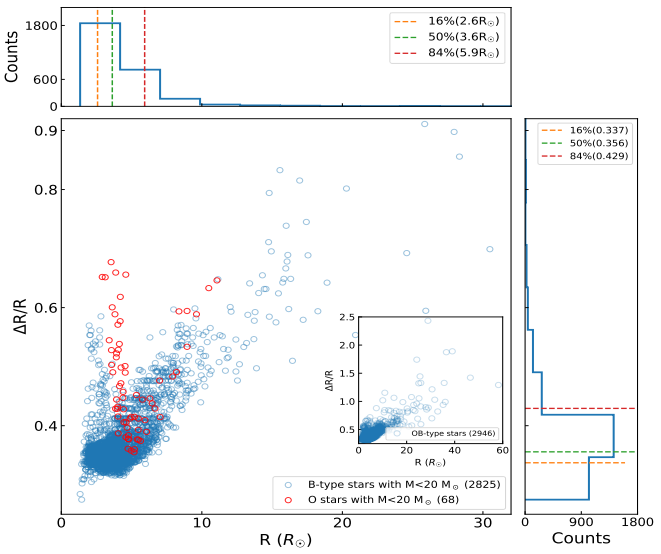}
    \caption{Similar to Figure \ref{fig:3} for the estimated radii and relative errors of OB-type stars in this work.}
    \label{fig:8}
\end{figure*}

\subsection{Luminosity and Radius}

Based on the stellar parameters from \citet{Guo2021}, the luminosities and radii of 2,946 OB-type stars are estimated by using the PARSEC isochrones and the relationship between luminosity and radius (L\,=\,4$\pi$\,R$^2$\,$\sigma$T$^4$), respectively. 

Figure \ref{fig:7} shows the distribution of OB-type stars in the luminosity (\,log\,(L/$\rm {L_\sun}$)) vs. luminosity relative error ($\Delta$\,L/L), the \,log\,(L/$\rm {L_\sun}$) of OB-type stars with masses less than 20\,M$_{\sun}$ ranges from 2 to 4. From the top histogram, it can be seen that the 84, 50, 16 percentiles of the distribution are less than 3.9, 3.2, 2.7, respectively. 
However, we also note that there is a larger luminosity error due to the relatively larger error of stellar parameters (the uncertainty of the surface gravity). Figure \ref{fig:8} shows the distribution of radii (R) and relative errors ($\Delta$\,R/R) of OB-type stars, and OB-type stars with masses less than 20\,M$_\sun$ are mainly located in the region of R\,$<$10\,R$_\sun$ and $\Delta$\,R/R\,$<$\,0.4. From the top histogram, we can see that the radii the distribution of 84\,$\%$ OB-type stars are less than 5.9\,R$_{\sun}$, the mean value of the radii are 3.6\,R$_{\sun}$, and there are only 16$\%$ stars with R\,$>$\,5.9\,R$_{\sun}$. 

In addition, we also derive the stellar photometry luminosity using photometric and astrometric data of $Gaia$ DR3 \citep{GaiaCollaboration2020}, based on bellow bolometric magnitude.

\begin{equation}\label{equation2}
\left\{
    \begin{aligned}
         log(L) = 0.4(M_{b,{\sun}}-M_b) \\
         \rm M_b = M'_G-A_G+BC_G \\ 
          M'_G = G+5+5log_{10}d\\ 
    \end{aligned}
\right.    
\end{equation}

where M$_{b,{\sun}}$ is the absolute bolometric magnitude of the Sun. The M$_b$ represents stellar absolute bolometric magnitude. The $G$ is the $G$-band magnitude, and the $d$ is the distance in pc from Early Gaia Data Release (EDR3) Distance \citep{Bailer-Jones2021}. The extinction A$_G$ is given by the blue-edge method \citep{Wang2014, Xue2016, Jian2017, Zhao2018, Zhao2020}. The bolometric corrections of stars are obtained from \citet{Jordi2010} and \citet{Bessell1998}, respectively. We have also listed the luminosities in Table \ref{chartable}.

\subsection{The Mass-Luminosity Relation}

\begin{figure*}
\xdef\xfigwd{\textwidth}
    \centering
     \includegraphics[width=1.0\linewidth]{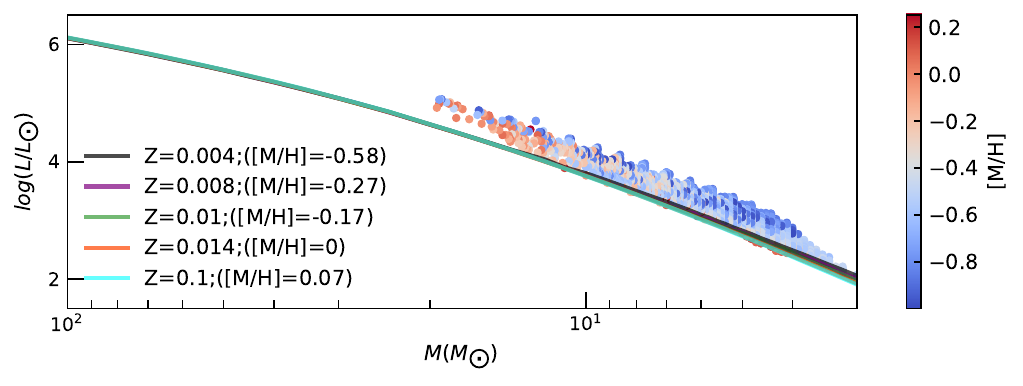}\\
	  \caption{The distribution of $M-L$ at different metal abundance, and the different color solid lines show the ZAMS line of PARSEC at each metal abundance. The dots mark the sample of normal B-type stars selected from the LAMOST LRS. The sample included 2,825 B-type stars.}
	  \label{fig:9} 
\end{figure*}

Figure \ref{fig:9} shows the distribution of our 2,825 B-type stars (M\,$\leq$\,20\,M$_{\sun}$) in the mass-luminosity ($M-L$) diagram. In Figure \ref{fig:9}, the different metallicity ([M/H]) zero-age main sequence (ZAMS) lines from the PARSEC model are marked using the different color solid lines, and the dots represent the 2,825 B-type stars. The color code is the metallicity of the stars. The almost independent between [M/H] and the $M-L$ relation is also consistent with the stellar model shown by the ZAMS lines.

According to the polynomial model of the $M-L$ -[M/H] relation from \cite{Xiong2022}, we also find the best-fit coefficients for this model using our OB-type stars. The polynomial model is written as:

\begin{equation}\label{equation3}
    log(L) =
    \!\begin{aligned}[t]
     & a_1+a_2log(M)+a_3log(M^2) \\
     &-0.755log(M^3)+0.189log(M^4)+a_4Z
    \end{aligned}
\end{equation}

where the values of a1, a2, a3, and a4 are -0.06, 4.399, 0.203, and -0.293, respectively. We use the polynomial of the model to fit the coefficients of the third- and fourth-order terms of $logM$, and show in Figure \ref{fig:9}. 

\section{Conclusions}\label{Conclusions}

In this work, we present the measurements of masses, luminosities, and radii of 2,946 OB-type stars in LAMOST DR5, based on the Padova isochrones and stellar parameters (effective temperature, surface gravity, and metallicity) from \citet{Guo2021}. 

Our analysis indicates that our method can provide a reliable masses estimation for OB-type stars with M\,$\leq$20\,M$_{\sun}$, based on their stellar parameters and the Padova isochrones model. We also note that it is difficult to estimate the accurate ages of OB-type stars due to their unknown evolution stages. The typical error of masses estimated is 21.4 percent for OB stars with S/N$_{\rm g}$ of spectra higher than 40. There is a good agreement between the distribution of their masses and their spatial positions. However, the problem of binarity affecting the mass determination of massive stars cannot be ignored, and more stellar observation information is needed to solve it.

For most OB-type stars, the luminosities obtained using our method are consistent with those calculated by using the data from $Gaia$ DR3. Moreover, we also fit the mass-luminosity relation of B-type stars by using our masses and the polynomial model of the $M-L$-[M/H] relation from \cite{Xiong2022}. 

\begin{acknowledgments}
We thank the anonymous reviewers for their helpful suggestions to help improve this manuscript. We thank Dr. Jianping Xiong for discussions of MLR and data support. This study is supported by the National Key Basic R$\&$D Program of China No. 2019YFA0405500, the National Natural Science Foundation of China under grants No.12173013, the project of Hebei provincial department of science and technology under the grant number 226Z7604G, and the Hebei NSFC (No. A2021205006, A2021205001). This study is supported by the Hebei Province Graduate Innovation Funding Project (CXZZBS2023086). We acknowledge the science research grants from the China Manned Space Project. Funding for the project has been provided by the National Development and Reform Commission. This work has made use of data from the  European Space Agency (ESA) mission Gaia (\url{https://www.cosmos.esa.int/gaia}), processed by the Gaia Data Processing and Analysis Consortium (DPAC, \url{https://www.cosmos.esa.int/web/gaia/dpac/consortium}). Funding for the DPAC has been provided by national institutions, in particular, the institutions participating in the Gaia Multilateral Agreement. The Guoshoujing Telescope (the Large Sky Area Multi-Object Fiber Spectroscopic Telescope LAMOST) is a National Major Scientific Project built by the Chinese Academy of Sciences. LAMOST is operated and managed by the National Astronomical Observatories, Chinese Academy of Sciences. 

\end{acknowledgments}

\appendix
\section{The measurement of stellar mass and age in the MCMC}

\begin{figure}
    \centering
    \includegraphics[width=16cm,height=16cm]{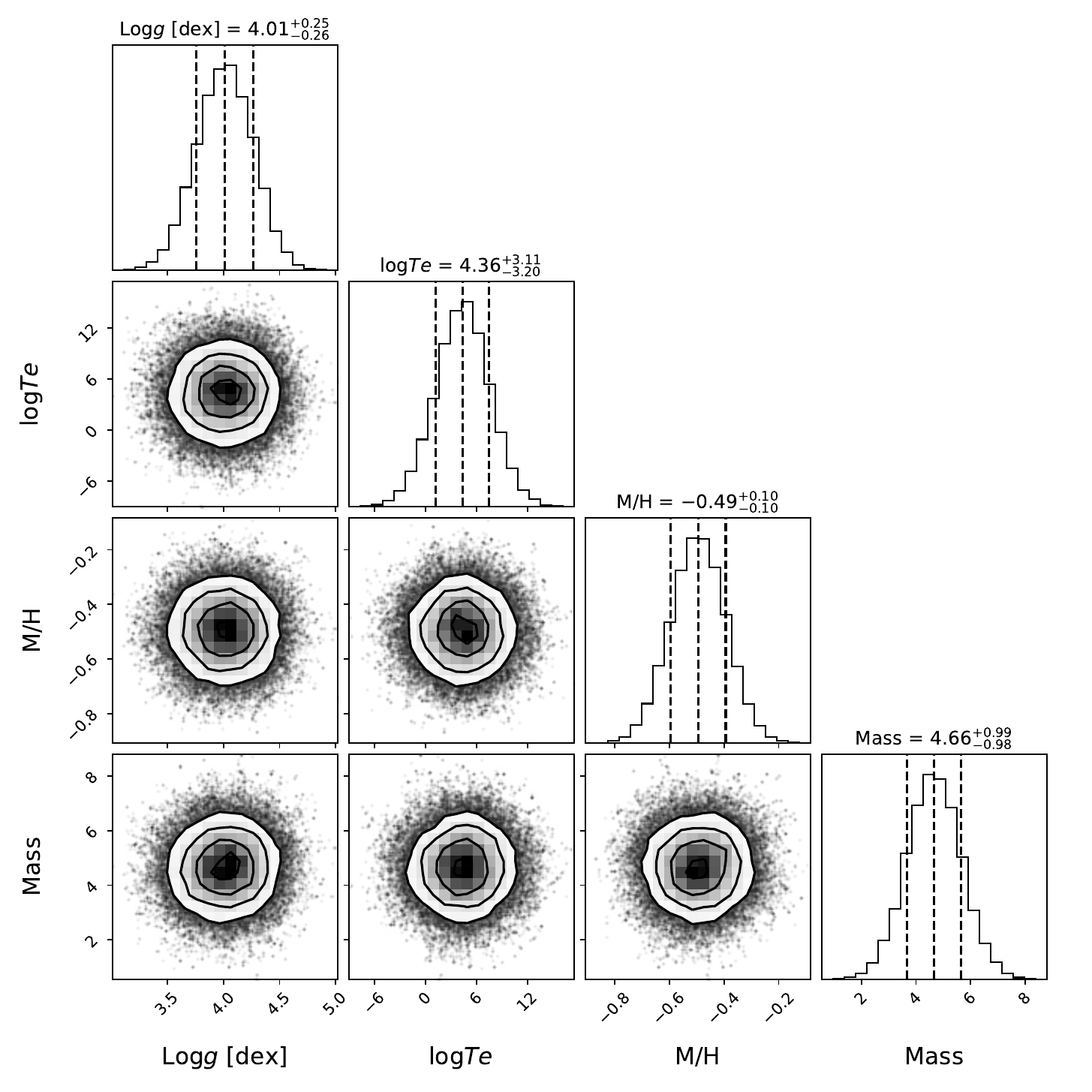}
    \caption{Predicting the posterior distribution of mass. The contours from inside to outside enclose 68{\%}, 95{\%}, and 99{\%} of the total probability in three non-diagonal panels. The three vertical dashed lines in the five diagonal panels represent the lower, median, and upper errors of the distribution at the 16, 50, and 84 percentile, respectively.\label{fig:10}}
\end{figure}

\bibliographystyle{aasjournal}
\bibliography{ref-lamost}

\begin{longrotatetable}
\begin{deluxetable*}{cccccccccccccccc}
\tablecaption{The basic information (RA, Dec) and parameters (mass, age, luminosity, and radius) of 2,946 OB-type stars in LAMOST.\label{chartable}}
\tablewidth{700pt}
\tabletypesize{\tiny}
\tablehead{
\colhead{Destigation} & \colhead{RA} & \colhead{Dec} & \colhead{S/N} & \colhead{$T_{\rm eff}$} & \colhead{log\,$g$} & \colhead{$[M/H]$} & \multicolumn{5}{c}{This work} & \multicolumn{3}{c}{Literature} & \\ 
\cline{8-12} &\cline{12-15}
\colhead{(LAMOST)} & \colhead{} & \colhead{} & \colhead{($g$-band)} & \colhead{(kK)} & \colhead{(dex)} & \colhead{}& M(\,M$_{\sun}$)& M$_2$(\,M$_{\sun}$) & log(\,L/L$_{\sun}$) & R(\,R$_{\sun}$) & log(\,L$_{Mbol}$/L$_{\sun}$) & M(\,M$_{\sun}$) & log(\,L/L$_{\sun}$) & R(\,R$_{\sun}$) & ref.\\
} 
\startdata
J005052.11+434533.5 &  12.717148 & 43.759318 & 329.71 & 18.2 &  4.1 &-0.3 & 5.3±1.05 & 5.6±1.15 & 3.01±0.28 & 2.98±1.49 & 2.91 & \nodata & \nodata & \nodata\\
J011021.32+574403.7 &  17.588851 & 57.734386 & 77.68 & 15.4 &  4.3 &-0.5 & 3.5±0.75 & 4.2±0.8 & 2.42±0.31 & 2.1±1.11 & 2.53 & \nodata & \nodata & \nodata\\
J013314.12+485727.9 &  23.30885 & 48.957763 & 64.05 & 17.6 &  4.2 &-0.4 & 4.6±0.95 & 5.0±1.0 & 2.78±0.3 & 2.46±1.29 & \nodata & \nodata & \nodata & \nodata\\
J013844.73+375509.9 &  24.686416 & 37.919419 & 56.16 & 16.1 &  4.0 &-0.6 & 4.3±0.95 & 5.0±1.05 & 2.86±0.32 & 3.21±1.73 & \nodata & \nodata & \nodata & \nodata\\
J015443.87+550657.5 &  28.683453 & 55.11589 & 390.46 & 17.3 &  4.4 &-0.2 & 4.4±0.95 & 4.6±0.9 & 2.6±0.31 & 2.05±1.0 & 3.11 & \nodata & \nodata & \nodata\\
J015749.01+562804.6 &  29.45537 & 56.467441 & 252.26 & 26.1 &  4.4 &-0.1 & 8.5±2.6 & 8.9±2.7 & 3.57±0.44 & 2.76±1.79 & 3.97 & \nodata & \nodata & \nodata\\
J015908.99+561810.3 &  29.787485 & 56.302871 & 219.02 & 21.5 &  4.4 &-0.2 & 5.9±1.5 & 6.2±1.5 & 3.04±0.36 & 2.21±1.3 & 3.21 & \nodata & \nodata & \nodata\\
J020040.30+542843.4 &  30.168527 & 54.479229 & 423.74 & 26.8 &  4.4 &-0.6 & 8.8±2.5 & 9.0±2.8 & 3.67±0.41 & 2.93±1.58 & 3.89 & \nodata & \nodata & \nodata\\
J020439.90+541617.0 &  31.166291 & 54.271391 & 380.15 & 17.5 &  4.1 &-0.6 & 4.7±1.0 & 5.4±1.05 & 2.95±0.3 & 3.01±1.54 & \nodata & \nodata & \nodata & \nodata\\
J021414.87+543928.8 &  33.561988 & 54.65802 & 282.07 & 17.8 &  4.2 &-0.6 & 4.6±1.0 & 5.4±1.1 & 2.86±0.31 & 2.61±1.39 & \nodata & \nodata & \nodata & \nodata\\
J022349.58+592918.2 &  35.9566 & 59.4884 & 57.67 & 15.6 &  4.2 &0.1 & 4.5±0.9 & 4.4±0.9 & 2.64±0.29 & 2.65±1.37 & 2.7 & \nodata & \nodata & \nodata\\
J022510.24+593332.4 &  36.2927 & 59.559 & 49.8 & 20.0 &  4.1 &-0.6 & 6.0±1.2 & 6.8±1.3 & 3.28±0.29 & 3.34±1.61 & 2.88 & \nodata & \nodata & \nodata\\
J022818.61+584949.5 &  37.0784 & 58.8309 & 43.16 & 17.9 &  4.2 &-0.5 & 4.8±1.0 & 5.4±1.05 & 2.87±0.3 & 2.6±1.36 & 3.01 & \nodata & \nodata & \nodata\\
J025340.52+565922.2 &  43.418845 & 56.989504 & 82.0 & 23.0 &  3.9 &-0.3 & 9.1±1.9 & 9.6±1.95 & 3.91±0.3 & 5.29±2.77 & 3.69 & \nodata & \nodata & \nodata\\
J030344.23+513940.9 &  45.934324 & 51.661382 & 101.16 & 27.1 &  3.9 &-0.8 & 11.8±2.85 & 13.5±2.7 & 4.33±0.35 & 6.17±2.9 & \nodata & \nodata & \nodata & \nodata\\
J032451.50+585622.7 &  51.214597 & 58.939656 & 54.14 & 21.8 &  3.9 &-0.3 & 8.0±1.7 & 8.5±1.8 & 3.74±0.3 & 4.84±2.46 & 3.26 & \nodata & \nodata & \nodata\\
J034840.79+561837.0 &  57.169989 & 56.310302 & 138.96 & 25.7 &  3.6 &-0.8 & 13.0±4.5 & 14.3±4.45 & 4.52±0.5 & 8.52±6.13 & \nodata & \nodata & \nodata & \nodata\\
J041110.86+504229.5 &  62.794727 & 50.708221 & 81.85 & 31.0 &  3.7 &-0.5 & 19.0±6.85 & 20.4±7.1 & 4.89±0.52 & 8.95±5.51 & \nodata & \nodata & \nodata & \nodata\\
J042911.09+534740.5 &  67.296226 & 53.794609 & 207.25 & 21.0 &  3.7 &-0.1 & 8.8±2.1 & 9.1±2.2 & 3.95±0.34 & 6.62±3.28 & \nodata & \nodata & \nodata & \nodata\\
J043001.38+531420.5 &  67.505776 & 53.239038 & 52.11 & 24.0 &  4.1 &-0.7 & 8.2±1.6 & 9.6±1.85 & 3.74±0.28 & 3.99±1.91 & \nodata & \nodata & \nodata & \nodata\\
J043638.52+543252.4 &  69.160513 & 54.547911 & 48.37 & 21.9 &  3.7 &-0.1 & 9.4±2.15 & 9.6±2.05 & 4.0±0.33 & 6.4±3.08 & \nodata & \nodata & \nodata & \nodata\\
J051118.11+393941.7 &  77.825478 & 39.661605 & 148.18 & 23.3 &  3.5 &-0.4 & 12.1±3.4 & 12.4±3.55 & 4.45±0.4 & 9.51±4.99 & \nodata & \nodata & \nodata & \nodata\\
J051833.69+423747.7 &  79.640411 & 42.629932 & 65.92 & 26.3 &  4.0 &-0.9 & 10.2±2.25 & 12.0±2.5 & 4.12±0.32 & 5.11±2.56 & \nodata & \nodata & \nodata & \nodata\\
J052512.84+361921.3 &  81.303522 & 36.322588 & 98.54 & 29.2 &  4.1 &-0.7 & 12.1±2.65 & 14.0±3.45 & 4.24±0.31 & 4.74±2.28 & 3.85 & \nodata & \nodata & \nodata\\
J060127.89+230828.1 &  90.36692 & 23.14098 & 453.81 & 16.7 &  4.0 &-0.2 & 5.1±1.05 & 5.2±1.05 & 2.99±0.29 & 3.47±1.83 & \nodata & 4.95±0.0 & 2.8±0.0 & 3.5±0.0&  [1],[2],[4]\\
J061059.16+115937.4 &  92.746537 & 11.993731 & 999.0 & 20.2 &  4.0 &-0.4 & 6.6±1.4 & 7.3±1.4 & 3.47±0.3 & 4.08±2.17 & \nodata & 5.2±0.03 & \nodata & \nodata&[1],[4],[5]\\
J061322.18+230920.1 &  93.342417 & 23.155605 & 269.32 & 18.5 &  4.1 &-0.6 & 5.1±1.05 & 5.8±1.15 & 3.04±0.29 & 2.98±1.5 & \nodata & \nodata & \nodata & \nodata\\
J061825.41+233415.3 &  94.6067 & 23.57145 & 419.05 & 28.3 &  4.1 &-0.2 & 12.1±2.85 & 12.6±2.85 & 4.15±0.34 & 4.6±2.28 & \nodata & 11.7±0.0 & \nodata & 5.6±0.0&[8]\\
J062522.62+195451.4 &  96.344258 & 19.914298 & 67.34 & 22.4 &  4.0 &-0.4 & 8.0±1.7 & 8.8±1.85 & 3.73±0.3 & 4.51±2.24 & 3.3 & \nodata & \nodata & \nodata\\
J062609.32+195026.9 &  96.538866 & 19.840814 & 222.24 & 20.4 &  4.0 &-0.4 & 6.7±1.35 & 7.4±1.4 & 3.48±0.29 & 4.05±2.1 & 3.23 & 6.407±0.039 & 3.127±0.132 & 3.662±0.136&  [6]\\
J063543.80+024240.4 &  98.93253 & 2.711233 & 453.59 & 30.9 &  3.7 &-0.4 & 19.6±7.0 & 21.6±7.6 & 4.95±0.51 & 9.62±5.61 & \nodata & \nodata & \nodata & \nodata\\
J063801.76+095300.8 &  99.507361 & 9.883571 & 449.31 & 17.3 &  3.8 &-0.5 & 5.4±1.3 & 6.2±1.4 & 3.29±0.34 & 4.54±2.53 & \nodata & 5.01±0.31 & \nodata & \nodata& [3]\\
J075738.35-071122.1 &  119.40983 & -7.189494 & 124.98 & 18.3 &  4.0 &-0.7 & 5.2±1.05 & 6.0±1.2 & 3.12±0.29 & 3.33±1.7 & \nodata & 4.95±0.0 & 2.81±0.0 & 3.54±0.0& [1],[2],[4]\\
J094844.71-024248.9 &  147.185359 & -2.713604 & 90.07 & 20.4 &  3.5 &-0.2 & 9.2±2.45 & 9.5±2.25 & 4.04±0.38 & 7.8±4.52 & \nodata & 8.55±0.53 & \nodata & \nodata& [3]\\
J151811.89+313849.2 &  229.547236 & 31.647009 & 96.94 & 15.2 &  4.2 &-0.5 & 3.6±0.75 & 4.2±0.85 & 2.45±0.3 & 2.26±1.21 & 2.65 & 4.47±0.28 & 2.41±0.2 & 2.79±0.11&  [1],[2],[4]\\
J192736.32+414205.5 &  291.901351 & 41.701539 & 421.42 & 17.8 &  3.9 &-0.2 & 5.9±1.2 & 6.2±1.0 & 3.28±0.29 & 4.21±2.19 & 3.31 & 2.37±0.0 & \nodata & 3.05±0.0&  [7]\\
...&...&...&...&...&...&...&...&...&...&...&...&...&...&...   
\enddata
    \tablecomments{(1). The stellar parameters ($T_{\rm eff}$, log\,$g$ and $[M/H]$)
    are estimated by \citet{Guo2021}.\\
    (2). We also provide log(\,L$_{Mbol}$/L$_{\sun}$) for stars with measured trigonometric parallaxes.\\
    (3). 
    $[1]$ \citet{Eker2014} \\
    $[2]$ \citet{Malkov2020} \\
    $[3]$ \citet{Hohle2010} \\
    $[4]$ \citet{Eker2015} \\
    $[5]$ \citet{Eker2018} \\
    $[6]$ \citet{Xiong2022}\\ 
    $[7]$ \citet{Surkova2004}\\
    $[8]$ \citet{Perevozkina1999}\\
    (This table is available in its entirety in machine-readable form.)
    }
\end{deluxetable*}
\end{longrotatetable}
\end{CJK}
\end{document}